\documentclass[numbers]{article}
\usepackage{arxiv}

\usepackage[utf8]{inputenc} 
\usepackage[T1]{fontenc}    
\usepackage{hyperref}       
\usepackage{url}            
\usepackage{booktabs}       
\usepackage{amsfonts}       
\usepackage{nicefrac}       
\usepackage{microtype}      
\usepackage{graphicx}
\usepackage{natbib}
\usepackage{doi}
\usepackage{array,tabularx}
\usepackage{setspace}
\hypersetup{hidelinks}

\title{Population Neuroscience 3.0\\
\large Bridging large-N and small-N paradigms in brain-wide association studies}

\date{Oct 22, 2026}	

\author{
    \href{https://orcid.org/0000-0002-9894-7934}{\includegraphics[scale=0.06]{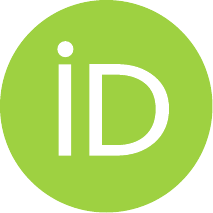}\hspace{1mm}Zi-Xuan~Zhou}
    \\
    State Key Laboratory of Cognitive Neuroscience and Learning, \\
    Beijing Normal University, Beijing 100875, China \\
    Developmental Population Neuroscience Research Center, \\
    IDG/McGovern Institute for Brain Research, \\
    Beijing Normal University, Beijing 100875, China \\
    \And
    \href{https://orcid.org/0000-0001-9110-585X}{\includegraphics[scale=0.06]{orcid.pdf}\hspace{1mm}Xi-Nian~Zuo}\thanks{Correspondence to Xi-Nian Zuo (\texttt{xinian.zuo@bnu.edu.cn})} \\
    State Key Laboratory of Cognitive Neuroscience and Learning, \\
    Beijing Normal University, Beijing 100875, China \\
    Developmental Population Neuroscience Research Center, \\
    IDG/McGovern Institute for Brain Research, \\
    Beijing Normal University, Beijing 100875, China \\
    National Basic Science Data Center, Beijing 100190, China \\
}

\renewcommand{\shorttitle}{Population Neuroscience 3.0}

\hypersetup{
pdftitle={Population Neuroscience 3.0},
pdfsubject={Neurons and Cognition (q-bio.NC); Quantitative Methods (q-bio.QM); Methodology (stat.ME)},
pdfauthor={Zi-Xuan~Zhou, Xi-Nian~Zuo},
pdfkeywords={Human neuroscience, Cognitive neuroscience, Population neuroscience, Brain--behavior associations, Brain charts},
}

\begin{document}
\maketitle

\begin{abstract}
\textbf{Recent advances and reflections on reproducible human neuroscience, especially brain-wide association studies (BWAS) leveraging large datasets, have led to divergent and sometimes opposing views on research practices and priorities. The debates span multiple dimensions, including large-N vs.\ small-N sampling, cross-sectional vs.\ longitudinal, observational vs.\ interventional/causal, population generalizability vs.\ individual specificity, data-driven vs.\ hypothesis-driven, prediction-oriented vs.\ explanation-oriented, and molecular/cellular vs.\ systems-level. Taken together, these tensions are not only about methodology, but also about how large-N population studies and small-N, mechanism-focused studies should relate to one another. Shifts along these axes have sharpened disagreements and added to fragmentation in an already heterogeneous field of cognitive neuroscience. Here, we sketch an integrative response grounded in population neuroscience, organized around a closed-loop ``design--analysis--interpretation'' research cycle that aims to build consensus while bridging these divides. Our central claim is that population neuroscience offers a distinctive population-level vantage point for identifying general patterns, characterizing inter-individual variabilities, and benchmarking intra-individual changes, thereby providing reference frameworks through which small-N, mechanism-focused studies can co-evolve with large-N population studies. Population neuroscience is not simply about providing larger N for BWAS; more fundamentally, it is about accumulating cross-scale priors and shared infrastructures for design, analysis, and interpretation across human neuroscience studies. In this sense, we outline a ``third-generation'' agenda for population neuroscience that shifts emphasis from amassing isolated associations to building reference frameworks for cumulative mechanistic and translational research.}
\end{abstract}


\section*{The Rise of Population Neuroscience}

In 2010, Tom{\'a}{\v{s}} Paus envisioned the benefits of combining genetics and epidemiology with cognitive neuroscience. This cross-fertilization would make it possible to study and understand how genetic and environmental factors shape the processes that lead to particular brain states and give rise to individual variability in the human brain, an agenda he termed \textit{population neuroscience} \cite{paus2010population}. In 2013, Falk \textit{et al.} \cite{falk2013representative} further articulated population neuroscience as a discipline at the intersection of population science and neuroscience, and elaborated its advantages: using tools from population science to recruit representative samples for neuroscience studies, and using neuroscience to open the neural ``black box'' in population research. They also emphasized the profound influence of context and culture on human behavior, and warned against assuming uniform brain--behavior relationships while ignoring these factors.

Greene \textit{et al.} \cite{greene2022brain} provided empirical support for these concerns in 2022. Using a widely adopted predictive modeling protocol in cognitive neuroscience \cite{shen2017using}, they showed that relationships between brain network patterns and behavioral traits differ systematically across subgroups defined by sociodemographic characteristics. This study confirmed Falk \textit{et al.}'s \cite{falk2013representative} earlier argument that population science is needed to reveal the complexity of brain--behavior associations. The accompanying editorial \cite{EDITORIAL2022cognitive} perceptively noted that the study by Greene \textit{et al.} \cite{greene2022brain}, together with the paper by Marek \textit{et al.} \cite{marek2022reproducible} on the sample size requirements of brain-wide association studies (BWAS) published five months earlier, marked a crossroads for research practices in cognitive neuroscience.

Since then, the term ``population neuroscience'' has been invoked with increasing frequency in two senses: (1) large-N studies that aim for more representative samples, and (2) BWAS that explicitly incorporate demographic, social, and environmental variables alongside behavioral measures \cite{gard2023weight,schumann2023addressing,hatzenbuehler2024research,ramduny2025representing,osayande2025quantifying,shaaban2026population}. At the same time, critical reflections on cognitive neuroscience have also fueled a countervailing movement that encourages research practices at the opposite end of several spectra: hypothesis-driven designs with small-N samples, task-based paradigms, experimental interventions, and dense longitudinal sampling that target individual-specific features \cite{gratton2022brain}. These debates have also revived long-standing disputes about the assumptions and approaches of brain research, for example over localization versus distributed accounts of brain function \cite{westlin2023improving}. Together, these currents have amplified tensions across perspectives, methods, and findings, and risk further entrenching divisions within the field.

Here, after briefly reviewing these divergences, we develop a reciprocal framework linking large-N population studies with small-N focused studies. Population studies can build reusable reference resources that situate focused observations within broader population variation; focused studies can then use deeper phenotyping, dense longitudinal sampling, and targeted experiments to examine questions that broad cohorts cannot resolve, while returning hypotheses for independent population-level evaluation. We organize this exchange as a design–analysis–interpretation cycle (Figure~\ref{fig:fig1}). Through this cycle, evidence from the two approaches can accumulate across studies and scales.

\section*{Recent Progress in Population Neuroscience}\label{sec2}
Since population neuroscience was first articulated, openly shared neuroimaging data have increased by orders of magnitude \cite{glasser2016human,miller2016multimodal,saragosa2022practical,ge2023increasing}, and this growth in data resources continues. With the volume of in vivo brain imaging now available, cognitive neuroscientists can tackle questions that were previously out of reach. These large-scale studies have also revealed important limitations of common practices in the field.

A particularly influential example, introduced earlier, is the study by Marek \textit{et al.} \cite{marek2022reproducible}. Using a total of approximately 50,000 participants from three datasets, they showed that the effect sizes of cross-sectional brain--behavior associations are much smaller than those commonly reported in the previous literature. This finding triggered extensive discussion and debate \cite{gratton2022brain,rosenberg2022establish,EDITORIAL2022revisiting,bandettini2022challenge,spisak2023multivariate,tervo2023reply,marek2025replicability,burns2025bias,uddin2025task}, prompting reflection on issues such as rethinking study designs \cite{gratton2022brain,rosenberg2022establish,bandettini2022challenge,kang2024study,ooi2025longer}, optimizing analysis methods \cite{rosenberg2022establish,bandettini2022challenge}, and integrating multimodal data \cite{song2022linking,schulz2024performance,tetereva2025improving,grosenick2026multimodal}. It also sharpened distinctions in research priorities, including whether to prioritize improving measurement reliability \cite{rosenberg2022establish,gell2024measurement}, the relative merits and complementary uses of resting-state and task-induced data \cite{rosenberg2022establish,marek2025replicability,uddin2025task,nau2024centering,luppi2026unlocking}, and renewed scrutiny of localization assumptions \cite{westlin2023improving}.

Among these reflections, the contrast drawn by Gratton \textit{et al.} \cite{gratton2022brain} between \textit{consortium studies} and \textit{focused studies} has been especially influential. Consortium studies typically emphasize large-N cross-sectional designs. Focused studies, by contrast, rely on small-N designs that are theoretically driven and tightly controlled, emphasize within-individual (longitudinal or repeated-measures) sampling, and seek large effect sizes induced by experimental interventions, development, or injury. Gratton \textit{et al.} also argued that the latter path may be closer to clinical translation and more accessible to early-career researchers who need projects that can be led and completed independently. This dichotomy has been reinforced by the growing literature on individual-level precision imaging \cite{michon2022person} and by debates about (group-level) nomothetic versus (individual-level) idiographic approaches \cite{mattoni2025group,mistry2026nonergodicity}. We refer readers to a more recent review by Gratton and colleagues \cite{lee2025using}, which discusses how these two paths can be integrated, an issue that is also central to the present article.

Another prominent population neuroscience study, published around the same time as Marek \textit{et al.} \cite{marek2022reproducible}, is the work by Bethlehem \textit{et al.} \cite{bethlehem2022brain}. Aggregating magnetic resonance imaging (MRI) scans from more than 100,000 participants, they modeled ``brain charts'' across the human lifespan, marking a major milestone for population neuroscience. Because such charts encode population-level information about both general patterns and individual variabilities, they can serve as ``microscopes'' that map data from raw measurement space into a more interpretable normative parameter space (see the conceptual demonstration below). Following this landmark paper, many studies have used large, population-based, age-diverse neuroimaging datasets to construct brain charts and to refine our understanding of fundamental principles in developmental neuroscience \cite{sun2025human,zhuo2025charting,sun2025population,kim2025white}, establishing a second major line of population neuroscience research alongside BWAS. In parallel, more large-scale developmental neuroimaging cohorts have been made openly available \cite{fan2023longitudinal,sadeghi2025interplay,shafiei2025reproducible}, providing critical infrastructure for this emerging brain-chart agenda.

\section*{Prospects for Population Neuroscience}\label{sec3}
In what follows, we place population neuroscience within the broader landscape of neuroscience and discuss the wider prospects opened by large-scale BWAS, brain charts, and related population-level agendas. Unlike earlier discussions that have mainly framed population neuroscience as a form of large-N neuroimaging aimed at improving sample representativeness, population generalizability, and the integration of demographic, social, and environmental factors into BWAS, we focus on population neuroscience as a long-term, systematic, and foundational ``slow-science'' effort \cite{kaelin2017publish,frith2020fast}. In this view, the goal is to build population-level reference infrastructures and shared priors that can constrain, compare, and gradually accumulate biological explanations, and to assemble a comprehensive human neuroscience picture across time scales, spatial scales, and analytic levels, thereby integrating currently fragmented findings. Viewed alongside earlier formulations by Paus \cite{paus2010population} and by Falk and colleagues \cite{falk2013representative}, this reference-framework perspective, centered on bridging large-N and small-N research, can be understood as a ``third generation'' agenda in population neuroscience.

\begin{figure*}[!t]
    \centering
    \includegraphics[width=\textwidth]{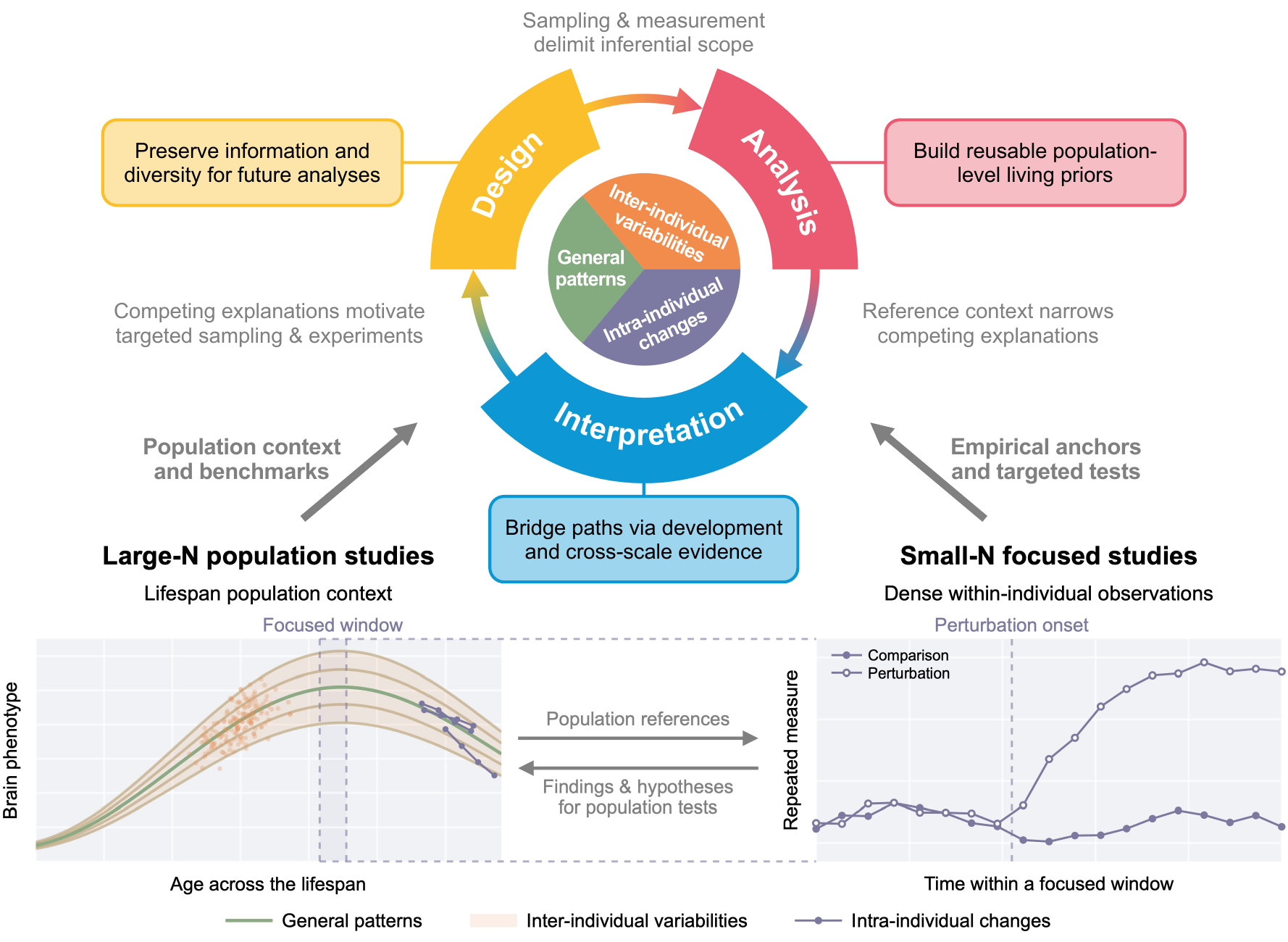}
    \caption{\textbf{Population references linking large-N and small-N research within a design--analysis--interpretation cycle.} The central cycle highlights the three levels of variation targeted by population neuroscience: general patterns, inter-individual variabilities, and intra-individual changes \cite{zhou2024population}. The lower-left schematic situates individual trajectories within lifespan population variation; the dashed window indicates a shorter timescale for focused studies. The lower-right schematic illustrates dense observations around a targeted perturbation. Population studies provide reference context and benchmarks, while focused studies generate findings and hypotheses for independent population-level evaluation and future study design. All trajectories and responses are schematic.}
    \label{fig:fig1}
\end{figure*}

As recent discussions have suggested \cite{lee2025using,gell2025psychiatric,ali2026balancing}, consortium studies and focused studies can be mutually reinforcing. Examining these two paths systematically from a population neuroscience perspective yields further insights: it clarifies how they can be linked, helps to mitigate the opposition between large-N and small-N research, and may ultimately reorganize how future studies are designed and linked. In addition, a population neuroscience framework can help bridge several long-standing divides, such as the relationship between brain structure and function, the nature--nurture debate, and the gap between molecular/cellular and systems-level knowledge in neuroscience. In the remainder of this section, we organize these prospects as a closed-loop ``research flywheel'' comprising three consecutive phases---design, analysis, and interpretation---and summarize this flywheel schematically in Figure~\ref{fig:fig1}.

\subsection*{Design}\label{sec3.1}
Current discussions of study design in population neuroscience have largely focused on improving the diversity and representativeness of recruited samples \cite{gard2023weight,kopal2023end,thompson2025ten}. This is undoubtedly a central issue: it directly inherits the methodological tradition of population science and addresses a dimension that was long neglected in cognitive neuroscience and neuroimaging. The importance of representative samples has been repeatedly demonstrated empirically, as different sampling frames yield systematically different brain--behavior associations \cite{greene2022brain,li2022cross,benkarim2022population,genon2022linking}, and in practice both selection and retention biases during recruitment, as well as exclusion decisions based on head motion \cite{ramduny2025representing,kay2025motion,mejia2026excessive} or other preprocessing criteria, can further erode representativeness. While acknowledging the importance of these issues, we shift the focus to several other design dimensions that have received less attention but are equally critical.

One useful way to organize the design space of population neuroscience is to view it as a multidimensional coordinate system \cite{gell2025psychiatric}. One axis is \textit{population coverage} (age, sex, ethnicity, socioeconomic status, cultural and geographic context, etc.). A second axis captures the \textit{temporal and sampling structure}, including timescale (single cross-section, multi-wave follow-up, intensive longitudinal designs, etc.) and multi-level sampling schemes. A third axis reflects \textit{measurement depth} (imaging modalities, task and naturalistic stimuli, genomics and multi-omics, behavioral and ecological measures, epidemiological and social-environmental variables, and so on). Cohorts such as the Generation R Study \cite{white2018paediatric}, the Adolescent Brain Cognitive Development (ABCD) Study \cite{saragosa2022practical,volkow2026abcd}, the HEALthy Brain and Child Development (HBCD) Study \cite{nelson2024introduction}, and the Chinese Child Brain Development (CCBD) project \cite{zhang2025diverse,liu2026landscaping} have made substantial progress across these axes through longitudinal follow-up, family information, multimodal imaging, and extensive phenotyping. Many questions nevertheless require denser repeated observations, additional sampling levels, or deeper measurement than broad cohorts can routinely provide. In what follows, we therefore focus on further development along the timescale/multi-level design and measurement depth axes, before returning to the overall design space and the ``bridging'' studies needed to address remaining gaps.

Cross-sectional designs and sparse longitudinal sampling remain important limitations: cohort effects and substantial inter-individual differences can bias estimates of intra-individual trajectories \cite{sorensen2021recipe,di2023mapping}. Without explicitly incorporating multiple longitudinal assessments into the design, it is difficult to treat development itself as an intrinsic dimension in BWAS. A typical example is the association between cortical morphology and intelligence quotient (IQ). Prior work suggests that this relationship is inherently dynamic and developmental, rather than a static cross-sectional correlation, and that adding longitudinal measurements reveals clear systematic divergence of cortical developmental trajectories between IQ groups \cite{shaw2006intellectual,schnack2015changes}. Similar design advantages have recently been demonstrated in a study of cortical function and IQ \cite{chen2025intellectual}. One of the core ideas in population neuroscience is to treat the developmental dimension as a key internal coordinate for explaining between-group differences at the population level. This point will be elaborated in the \textbf{Interpretation} section using Tinbergen's four questions as a multi-level explanatory framework \cite{tinbergen1963aims,bateson2013tinbergen,nesse2013tinbergen}.

A fundamental advantage of population neuroscience over traditional neuroimaging is that it introduces a population-level viewpoint. Long-term follow-up \cite{zhang2025dlpfc} of large samples at this level is, in effect, akin to running thousands of natural experiments (quasi-experiments) \cite{liu2021quantifying,bailey2024causal} with strong ecological validity. In human brain research, we cannot easily perform gene editing as in animal studies, cannot systematically dissect living brains, and are heavily constrained in the interventions we can ethically apply. Against this backdrop, the ability to observe large numbers of individuals as they naturally develop, age, and experience life events, with systematic and reasonably dense sampling, is especially valuable. With sufficiently long follow-up in large cohorts, the natural incidence of disease ensures that transitions from ``healthy'' to ``ill'' states will be captured \cite{miller2016multimodal}, often with multiple time points before and after onset. In such designs, naturally occurring transitions can be used to construct quasi-experimental comparisons, partially bridging the gap between observational and interventional research while preserving ecological validity. By analogy to the multi-level designs widely used in behavioral and imaging genetics, including twin studies \cite{friedman2021twin}, pedigree designs \cite{glahn2010genetic}, cross-cohort replication, and cross-ancestry analyses, population neuroscience cohorts should likewise incorporate family and twin samples, cohorts spanning different cultural and institutional contexts, and multi-level biological and social data across generations. This would allow us to trace when and how individuals diverge from common developmental trajectories, and to disentangle the respective contributions of genetic, shared environmental, and non-shared environmental factors.

Beyond the time dimension and multi-level sampling, measurement depth is another critical design axis in population neuroscience. Ideally, population studies would integrate as many relevant layers as feasible: multimodal imaging (structural, functional, diffusion-weighted, PET, EEG, MEG, etc.), a diverse set of task and naturalistic paradigms, genomic and multi-omic data, clinical and behavioral assessments, and ecological and epidemiological variables spanning family, community, policy, and cultural contexts, including the timing and co-occurrence of environmental exposures \cite{genon2026linking}. In practice, however, many cohorts remain shallow along this axis. Administrative and communication costs of cross-disciplinary collaboration, limits on the depth of team integration, and the challenges of maintaining harmonized protocols across sites all constrain what can be collected. As a result, key environmental variables that strongly influence brain and behavior are often missing or measured only coarsely. Even when tasks or questionnaires are collected, data quality may be suboptimal, and cross-site harmonization \cite{neidhart2024protocol} may be insufficient to support fine-grained analyses.

If population neuroscience is to deliver on its promise of detecting shortcut learning \cite{geirhos2020shortcut} and informing causal graph analyses, designs must broaden the coverage and resolution of covariates and environmental measures from the outset. For example, Marek \textit{et al.} \cite{marek2025replicability} highlighted sleep disturbance as a potential shortcut feature in models predicting psychopathology from functional connectivity, implying that cohorts must measure sleep, lifestyle, and related factors with sufficient precision. At the same time, many innovative task paradigms are currently confined to small laboratories and have not yet been incorporated into large consortia. Conversely, many ``classic'' tasks that have achieved broad consensus in cognitive science are precisely those with minimal between-individual variability, well-suited to detect robust group-level effects but poorly suited to characterize individual differences \cite{hedge2018reliability}, which are the very focus of population neuroscience. This reality further underscores the importance of sustained cross-disciplinary collaboration and of systematically importing the most informative paradigms from small-N precision studies back into cohort design \cite{tibon2022bridging}.

It is also important to recognize that neuroimaging and related measurement technologies themselves are evolving rapidly. Advances in acquisition hardware, reconstruction algorithms, preprocessing pipelines, and analysis methods are likely to continue in the near future. If only derived measures from current pipelines are retained, without comprehensive archiving of the underlying raw data (including pre-reconstruction raw imaging signals, original behavioral recordings, and key metadata), the valuable cohorts will be difficult to reanalyze using improved methods in the future. From a population neuroscience perspective, preserving raw data, documenting acquisition, reconstruction, calibration, and measurement uncertainty, and providing reusable analysis interfaces \cite{karakuzu2025rethinking} for future reanalysis and cross-study comparability are integral components of study design. In short, the complexity of analysis methods should not exceed the information content provided by the design; good designs should preserve degrees of freedom for future generations of analytic methods, rather than locking studies permanently into current pipelines.

Bringing these strands together, we can view sample representativeness and diversity (population coverage), temporal and sampling structure (timescale and multi-level design), and measurement depth as three basic axes of the design space in population neuroscience. Substantial opportunities remain to increase temporal resolution, enrich multi-level sampling, and deepen measurement. Addressing these gaps will require specifically designed ``bridging'' studies. These may include embedding high-density longitudinal sub-cohorts within representative samples; enriching deep phenotyping studies with additional imaging modalities, novel task paradigms, and more comprehensive genomic and multi-omic data; and coordinating multiple cohorts so that different projects complement one another across the three axes. Meanwhile, mechanism-focused studies often push much further along the other two axes, with small samples but very deep measurement and dense longitudinal sampling. Through data aggregation across laboratories, standardization efforts, and the sharing of reusable code, such small-N deep studies can gradually scale up in sample size and, together with population cohorts, fill out this design space. This again highlights the importance of open science practices: without standardized metadata, reusable preprocessing code, and open access to raw data, efficient cross-study aggregation and analysis will be impossible.

\subsection*{Analysis}\label{sec3.2}
Many current analyses of population neuroscience data still rely on simple univariate, linear methods, for example assessing associations between individual brain structural or functional measures and demographic, social, or environmental variables. As Gratton \textit{et al.} \cite{gratton2022brain} pointed out, as long as such associations are reliable, even small effect sizes can have far-reaching social impact by informing public health policy, much like the classic relationship between childhood lead exposure and reduced IQ \cite{searle2014tracing}. At the same time, as Bandettini \textit{et al.} \cite{bandettini2022challenge} emphasized, there remain many ``unknown unknowns'' about the brain and the mechanisms underlying neuroimaging signals. Changing preprocessing or analysis pipelines, introducing different covariates or confounders, or simply analyzing a different population or cohort can qualitatively alter observed brain--behavior associations, not to mention the effects that emerge when mediators and moderators along alternative causal pathways are considered. More complex multivariate and nonlinear methods, including black-box models such as deep learning \cite{schulz2024performance,grosenick2026multimodal,abrol2021deep}, have been explored in some studies, but these do not by themselves resolve the underlying fragility. These practices suggest that population neuroscience remains at an early stage of analytic development. Progress depends on sample size and quality and on research priorities: rapidly extracting publishable associations can crowd out investment in long-term analytic infrastructure. In the small-N era, debates over pipelines and models were often confined to particular datasets. Population-level data offer the scale and diversity needed to evaluate these choices systematically within a shared framework, provided that sustained effort is directed toward building that infrastructure.

In our view, population neuroscience should be understood as a form of slow science. Its primary mission is not to employ complicated models to derive as many significant results as possible, but to construct and consolidate a family of infrastructures that can be queried, reused, and updated over time. A crucial point is that many of the priors implicit in current analysis pipelines are methodological legacies from the small-sample era, derived from group-average analyses in relatively homogeneous samples of middle-class, White adults. This is true for widely used brain templates, segmentation and registration algorithms, and also for common practices in parcellation, spatial smoothing, and statistical modeling: many of the tools currently adopted in population-level studies of individual differences were in fact designed, iterated, and selected to better estimate an ``average brain,'' rather than to optimize the characterization of individual differences. When applied to more diverse populations, these tools may be unreliable or systematically biased: they can increase noise and struggle to capture meaningful inter-individual variability (while continuing to recover stable group averages). Calls from the precision neuroimaging literature \cite{michon2022person,lee2025using,hermosillo2024precision,dworetsky2024two,gratton2025dense} to reconsider traditional group-level analyses can be read as highlighting this problem from another angle: the basic algorithms needed for robust individual-differences analysis are still underdeveloped, and population-level datasets provide precisely the scale and diversity needed to build them. Table~\ref{tab:reference_resources} summarizes the reference resources considered below and their reciprocal roles in population and focused studies.

\begin{table*}[!t]
\centering
\caption{Reference resources linking population and focused studies.}
\label{tab:reference_resources}
\small
\setlength{\tabcolsep}{3pt}
\renewcommand{\arraystretch}{1.05}
\begingroup
\setstretch{0.95}
\begin{tabularx}{\textwidth}{@{}
>{\raggedright\arraybackslash}p{0.11\textwidth}
>{\hsize=1.10\hsize\linewidth=\hsize\raggedright\arraybackslash}X
>{\hsize=1.05\hsize\linewidth=\hsize\raggedright\arraybackslash}X
>{\hsize=0.85\hsize\linewidth=\hsize\raggedright\arraybackslash}X
@{}}
\toprule
\textbf{Resource family} &
\textbf{Reference content to build and maintain} &
\textbf{Use in focused studies} &
\textbf{Feedback guiding population studies} \\
\midrule

Methods and measurement references &
Templates and alignment models across diverse populations; preprocessing/QC standards; benchmarks of measurement reliability and bias. &
Select measurements and pipelines; align observations across studies while preserving individual differences. &
Identify measurement needs and method limitations for evaluation across populations. \\

\addlinespace[3pt]
Phenotype distributions and trajectories &
Phenotype distributions conditional on age and other relevant variables; shared reference axes; distributions of intra-individual change. &
Locate individuals and observed change within population variation; guide targeted recruitment and the timing of repeated measurements. &
Identify informative ages, timescales, and subgroups for future sampling. \\

\addlinespace[3pt]
Effect and cross-scale libraries &
Effect estimates with uncertainty; population association patterns; molecular, cellular, developmental, and functional annotations. &
Benchmark effect sizes for study planning; assess proposed specificity against population association patterns; prioritize candidate explanations using cross-scale annotations. &
Propose candidate pathways and subgroup-specific hypotheses for population-level testing. \\

\addlinespace[3pt]
Models, benchmarks, and simulations &
Reference confound models; foundation models; analytic method benchmarks; biophysical parameter distributions and virtual cohorts. &
Reuse reference models and learned representations; compare analytic choices; evaluate model predictions in virtual cohorts. &
Suggest model refinements and data needs for further development and independent evaluation. \\

\bottomrule
\end{tabularx}
\par\endgroup
\par\smallskip
\begin{minipage}{\textwidth}
\footnotesize\raggedright
These complementary resource families outline directions for building \textit{living priors}. Feedback guides subsequent population studies; proposed reference updates undergo independent evaluation. QC, quality control.
\end{minipage}
\end{table*}

A natural starting point for future analyses in population neuroscience is to construct base-layer priors from population-level data, particularly those directly related to preprocessing and quality control (QC). For example, we can build families of brain templates that better respect individual anatomical variability across age, sex, ancestry, and brain morphology \cite{ge2023increasing,dong2020charting,arthofer2024internally,guo2026motifs}, and then develop segmentation and registration algorithms that generalize across populations. A recent lifespan-generalizable skull-stripping model that leverages atlas-based priors \cite{wang2025lifespan} offers a clear case in point: when the new model was applied to the same population dataset in place of the old model, extreme outliers in estimated brain volume nearly disappeared, the lifespan trajectory of brain volume showed a more gradual decline in older age, and between-individual variability was substantially reduced at all ages. Together with improved skull-stripping accuracy, these changes suggest that a considerable fraction of the ``individual differences'' observed under the old model reflected systematic bias and noise rather than true variability. It is reasonable to expect that much more challenging steps such as cortical surface reconstruction \cite{lian2026unisurf}, fine-grained parcellation \cite{liu2026brainparc}, and cross-individual registration will benefit even more from algorithms trained on population-level data. Moreover, current approaches to aligning functional data across individuals generally depend on structural alignment, implicitly assuming that structure--function coupling is identical across individuals. This may be acceptable when the main goal is to recover group-average signals, but it is increasingly untenable in a field that explicitly foregrounds individual differences. Developing functional alignment methods that better respect individual differences, such as hyperalignment \cite{haxby2020hyperalignment, feilong2021neural, zhang2025boosting}, using large and diverse datasets is therefore a key direction at this level.

At the preprocessing and QC layer, pipeline choices themselves can also be viewed as a form of method prior: whether to apply global signal \cite{bolt2025autonomic} regression, which filtering and noise-modeling strategies to use, whether to operate at the voxel or parcel level and with which parcellation \cite{bryce2021brain}, how to set motion-censoring thresholds, and so forth. Large, diverse samples support systematic evaluation of pipeline performance and variance structure across populations, with measurement bias assessed against suitable validation benchmarks. These assessments can be shared as ``method maps'' and ``measurability maps.'' The former summarizes how different preprocessing choices perform in different analytic scenarios; the latter indicates which regions and measures are relatively more reliable, sensitive, and stable at different ages, in different ancestries, or across disease states. Such information can directly guide preprocessing and QC choices in small-sample, deep-phenotyping studies, including decisions about data sufficiency for individual mapping \cite{lynch2026objective}, and can also serve as a foundation for cross-cohort meta-analyses and normative model construction. This vision closely parallels well-established practices in genetics: before running a genome-wide association study, researchers adhere to a standardized QC pipeline (sample quality control, relatedness checks, population structure, imputation, single nucleotide polymorphism filtering, batch-effect control, etc.) to ensure that results from different cohorts can be combined and compared. By analogy, population neuroimaging both motivates and enables the development of comprehensive, reusable preprocessing/QC standards and documentation across projects \cite{koch2025versatile}. Only with deeper consensus at this foundational level can we hope to achieve more precise and larger-scale analytic syntheses.

On top of these base layers, one can then consider normative priors for brain and behavioral phenotypes themselves. Given that age is a robust and fundamental source of variance and a core focus of developmental neuroscience, the brain chart work discussed earlier represents a typical example of building phenotype priors along this dimension \cite{bethlehem2022brain,sun2025human,zhuo2025charting,sun2025population,kim2025white}. Leveraging large, representative samples across different life stages, brain charts can precisely track lifespan trajectories of diverse structural and functional measures. These trajectories refine our knowledge of general developmental patterns and, at the same time, provide a reference frame for characterizing inter-individual differences and intra-individual longitudinal changes. For example, population neuroimaging data may gradually support the construction of lifespan reference trajectories for functional gradients \cite{zhou2024lifespan,bethlehem2026brain}. At present, however, research has not yet sufficiently converged on group-level general patterns \cite{dong2021shifting,taylor2026functional,li2026spatiotemporal}. This likely reflects not only the intrinsic noisiness of functional MRI (fMRI), but also differences in sampling design (longitudinal versus cross-sectional), imaging protocols, the racial and cultural composition of populations, analytic choices \cite{dong2021shifting,taylor2026functional,li2026spatiotemporal,hong2020toward,byeon2026developmental}, and strategies for cross-individual gradient alignment \cite{sasse2026individual,bethlehem2026brain,bennani2010anisotropic}. As these priors are further iterated toward a more stable research consensus, precision fMRI or small-sample longitudinal studies can then perform finer parcellation and functional alignment for each participant along these age- and sex-specific reference axes, treating individual differences as local perturbations around the shared axes. In large cohorts with multiple longitudinal waves, we can go further and not only compute deviation scores at each time point, but also directly model the distribution of intra-individual change, thereby obtaining population-level priors on longitudinal dynamics. Taken together, brain charts and their generalized extensions provide a unified language and coordinate system for general patterns, individual differences, and longitudinal changes \cite{zhou2024population}, as discussed in detail in the conceptual demonstration below.

Furthermore, the most valuable output of BWAS and related population-level analyses is often not isolated statistically significant associations or phenotypes, but a high-dimensional prior over population variation. Such priors describe how brain phenotypes systematically shift with age, sex, socioeconomic status (SES) \cite{osayande2025quantifying,marek2025patterns}, and environmental exposures \cite{legaz2026exposome}; how the structure of variance itself changes; which patterns are highly conserved (low variance) and which are highly diverse (high variance). Recent ABCD analyses illustrate this broader structure of population variation: a common pattern across many BWAS maps was most reflective of SES \cite{marek2025patterns}. Extending such references across populations requires evaluating the comparability of multidimensional, context-dependent SES measures \cite{howe2012measuring}; measurement models may partly accommodate non-equivalence where suitable shared indicators are available \cite{curran2014moderated}. Beyond demographic and socioeconomic dimensions, integrating imaging and phenotypes across organs and modalities can further enrich these population-level references. Using multi-organ imaging \cite{mccracken2022multi,multi2025brain}, genetic data, and exposure measures, we can chart population-level networks linking the brain to the cardiovascular, metabolic, and immune systems \cite{villringer2025brain}, and thereby better understand the biological basis of comorbidity patterns (e.g., between depression and cardiovascular disease, or between obesity and cognitive decline). These patterns can provide ``boundary conditions'' and plausible ranges of effect sizes for fine-grained mechanistic work, and can also define data-driven brain--body--symptom axes: in small-sample intervention studies, we need not reinvent new dimensions from scratch, but can instead use axes learned from large populations \cite{skampardoni2026coupled} to characterize individuals' brain--body--behavior profiles and their shifts before and after treatment.

In parallel, population neuroscience can also systematically produce ``effect summary libraries'' for mechanism-focused research, analogous to genome-wide association studies (GWAS) summary statistics \cite{pasaniuc2017dissecting,dall2024psychiatric}. Publicly shared GWAS summary statistics make it possible to estimate single nucleotide polymorphism heritability and genetic correlations, perform multi-trait GWAS, fine-mapping, Mendelian randomization, and many other downstream analyses at the summary-data level; resources such as the GWAS Catalog \cite{cerezo2025nhgri} and PGS Catalog \cite{lambert2021polygenic,lambert2024enhancing} then collect and standardize these results, enabling reuse and comparison across cohorts. Building on resources such as the ENIGMA Toolbox \cite{lariviere2021enigma}, broader neuroimaging libraries could include BWAS effect matrices at parcel, network, or gradient levels; systematic virtual histology findings \cite{patel2021virtual} linking population-level maps of individual differences to ex vivo transcriptomics, receptor distributions, and cell-type markers; and derived quantities that support meta-analytic decoding and the construction of poly-phenotype scores. Such libraries can also provide empirically calibrated effect-size benchmarks for planning focused studies \cite{noble2026hidden}. Crucially, these high-dimensional priors are not only inputs for subsequent analyses; they are themselves high-value scientific results. For example, they may reveal that disorder-related patterns do not arbitrarily affect isolated patches of cortex, but tend to extend along pre-existing axes of population variation, thereby offering powerful constraints for mechanistic models and cross-scale integration.

Population neuroscience can also serve as a methodological platform \cite{dick2026adolescent}. On one front, population-level data can be used to model associations of commonly measured covariates with imaging or genetic phenotypes \cite{alfaro2021confound,radosavljevic2026controlling}, yielding reference confound models. These models can then provide effect priors (to be incorporated as fixed effects or constraints) and candidate covariate sets that smaller studies can adapt to their research questions and assumed causal structures, including when deciding whether to adjust for SES or other environmental factors \cite{vanderweele2019principles}. This reduces the need for each laboratory to choose covariates from scratch based on intuition. On another front, different analytic methods (e.g., dimensionality-reduction algorithms, network measures, decomposition and clustering approaches, causal discovery algorithms) can be systematically benchmarked \cite{liu2025benchmarking} on the same large dataset, creating an open ``method benchmark'' resource. This type of methodological meta-analysis is itself a foundational contribution: it can help researchers understand which methods are more robust and interpretable under which conditions. Population-level data can also be used to estimate biophysical parameters related to brain structure and function, together with their distributions in real populations, thereby providing more realistic priors for generative and simulation studies \cite{astle2023toward,liu2024generative,oldham2025coming,idesis2024generative,wang2024virtual} to construct virtual cohorts with age-, sex-, and subgroup-specific characteristics, serving as reference resources for modeling and methodological research, individualized inference, and precision intervention studies.

Moreover, population neuroscience bears the responsibility of building population-generalizable brain maps and models with uncertainty estimates, and of systematically annotating where evidence is sparse, so that subsequent clinical tools and machine learning (ML) systems do not inadvertently widen health disparities. A related line of work uses ML to build predictive models mapping high-dimensional brain phenotypes onto behavioral phenotypes in large samples \cite{tian2026spontaneous}. These models can have direct translational potential, but also face challenges of overfitting, shortcut learning, site effects, interpretability, and data leakage. For population neuroscience, an additional goal is to make models trained on large cohorts reusable in smaller studies \cite{he2022meta,wang2025big}. Brain-age models offer a simple example of a reusable output: the ``brain-age gap'' (predicted minus chronological age) can be used in smaller studies to investigate its biological and clinical relevance \cite{zhang2025brain,whitmore2025current}. Brain \textit{foundation models} broaden this principle of reuse by learning generic representations from large research and clinical imaging datasets for adaptation across downstream tasks \cite{dong2025brain,kondepudi2026health}. Such representations can improve predictive performance with limited task-specific labeled data \cite{tak2026generalizable} and can also serve as quantitative phenotypes for further research, as illustrated by SSDiff's diffusion MRI features used for phenotype prediction and genome-wide association analyses \cite{gong2026foundation}. Mechanistic interpretation requires explicit biological hypotheses about these representations and independent tests, for example whether a feature proposed to index plasticity tracks within-individual responses to a targeted perturbation. An important task for population neuroscience is therefore to train, evaluate, and maintain these models as shared resources for cumulative research across studies and datasets.

Finally, these population-derived resources should be maintained as \textit{living priors}: long-lived, reusable infrastructures that are evaluated and updated over time. All BWAS and downstream small-sample studies can then express their results within these shared frames (for example, by mapping diffusion measures to shared white-matter pathways \cite{li2026connectome}, or projecting onto a common gradient space or normative deviation space), enabling alignment and integration across projects and scales. Resources such as Neurosynth \cite{yarkoni2011large}, the ENIGMA Toolbox \cite{lariviere2021enigma}, and neuromaps \cite{markello2022neuromaps} represent important steps toward such cross-study integration and interpretation. Mirroring the versioning of reference genomes \cite{nurk2022complete,liao2023draft}, we can progressively build and maintain a family of versioned resources: reference brain charts, spatial templates, virtual histology atlases, poly-phenotype weights, and population ML feature libraries. Focused studies can pose local mechanistic questions within a reference version fixed before target analysis and contribute annotations, hypotheses, and design priorities for subsequent population studies. These population studies can support reference updates using larger and more diverse samples, with proposed changes evaluated on data held apart from their development and selection \cite{kriegeskorte2009circular}. Building on these efforts, the broader goal is to develop continually updated reference frameworks that integrate evidence across scales and modalities and characterize population trends, individual differences, and longitudinal changes using representative population data. The population-level trends and distributions of inter- and intra-individual variabilities captured in these infrastructures are themselves high-value scientific information: they serve both as priors and constraints for later analyses and modeling, and as key sources of inspiration for generating new hypotheses and designing new experiments. Schematically, population data $\rightarrow$ prior templates/models $\rightarrow$ mechanistic small-N studies $\rightarrow$ hypotheses and design priorities informing independent population studies and validated reference updates, forming a closed loop. The axis running through this closed-loop research flywheel is precisely these constantly updated living priors.

\subsection*{Interpretation}\label{sec3.3}
Within this research cycle, analysis and interpretation are tightly coupled. Analysis focuses on how data are processed and modeled at the mathematical and statistical levels, including how associations and uncertainty are quantified. Interpretation relates the assumptions and results of those computations to real-world physical, chemical, biological, and social entities and their relationships, embedding specific findings within a broader explanatory framework and extending the reasoning further. Within population neuroscience, interpretation also functions as a ``design document'' for the next round of research. It shapes decisions about what data to collect next, how to refine analyses, and at which levels and timescales to densify observation. In this section, we therefore consider interpretation in its own right, with two questions in mind. First, what specific opportunities does population neuroscience offer at the interpretive level, and how can these opportunities enable new research pathways and paradigms that bring different subfields of neuroscience into a unified framework? Second, how can interpretation build on the preceding phases of design and analysis and feed back into subsequent study design, thereby helping to close the loop of the research flywheel?

At its origin, population neuroscience is motivated by a core question: \textit{when} and \textit{how} do inter-individual differences in the human brain emerge from initially shared developmental trajectories? In other words, instead of treating individual differences as a static snapshot, population neuroscience views them as the accumulated outcome of different intra-individual trajectories over development \cite{paus2012some,zuo2018developmental,zuo2026developmental}. Recent precision neuroimaging research has shown, for example, that while the overall organization of brain networks is similar in children aged 8--12 years and in young adults, inter-individual variability is systematically lower in children, suggesting that individual differences in network organization may accumulate gradually across the lifespan \cite{demeter2025precision}. This provides empirical support for the idea that the brain follows an ``individual differentiation'' process. A population-level perspective aims to harness this variability: on the one hand, by systematically characterizing how genes and environmental factors shape the ontogeny of brain structure and intrinsic function; on the other hand, by using this knowledge to understand the combined nature--nurture pathways that drive developmental divergence, and by treating analyses of these divergence processes as a basis for unpacking brain complexity and approaching more fundamental questions about the human brain. In Tinbergen's four-question framework \cite{tinbergen1963aims,bateson2013tinbergen,nesse2013tinbergen}, ontogeny occupies a pivotal, intermediate position \cite{bergman2022leveling}. Upward, it connects to phylogenetic and functional explanations: principles of human brain development and individual differentiation can be compared with developmental patterns and adaptive mechanisms across species \cite{nowakowski2025new}, exploiting parallels between developmental and evolutionary processes for mutual constraint and insight. Downward, the time window can be narrowed to dense longitudinal neuroimaging combined with interventional paradigms to probe shorter-timescale proximate mechanisms \cite{ren2026parkinson,ren2026circuit,elliott2026precision}, especially during key phases such as the so-called ``mini-puberty'' in infancy, puberty, pregnancy, and menopause \cite{beck2023puberty,pritschet2024neuroanatomical,vinci2025dense}. This allows more fine-grained developmental pictures to be drawn and related to proximate mechanisms at even shorter time scales (e.g., from molecular and cellular biology) \cite{tsyporin2026competing}. Conducting multi-level, ecologically valid behavioral and neural investigations within this framework is likely to be a key driver of progress in population neuroscience.

At the level of interpretation, population neuroscience also offers opportunities for cross-scale evidence triangulation, which can help counter the ``fragmentation'' of contemporary neuroscience. Virtual histology provides one illustrative example. Patel and colleagues \cite{patel2021virtual}, working with large ENIGMA samples for six psychiatric disorders, conducted a meta-analysis of case--control cortical thickness differences across 34 regions and applied principal component analysis to extract a first principal component that explained approximately half of the variance in cross-disorder effects, interpreted as a shared spatial pattern of cortical thickness differences across diagnoses. They then aligned this macro-scale spatial axis with cell-specific gene expression maps and, using co-expression and enrichment analyses, identified gene modules associated with CA1 pyramidal cells, astrocytes, and microglia, with the pyramidal- and astrocyte-related modules enriched for disorder-associated genes across multiple psychiatric conditions. This provided candidate cellular and molecular clues for the observed population-level pattern. More generally, similar spatial axes can be systematically compared with a wider range of cellular and molecular maps to test for convergence at the level of cell types, receptor systems, and molecular pathways \cite{bazinet2023towards,lotter2024regional}, although the interpretive power of such comparisons remains constrained by the representativeness of the reference maps, the quality of cross-modal and cross-individual alignment, and repeated reuse of the same reference brain maps \cite{royer2026opportunities}. Mechanistic hypotheses are most convincing when supported by converging evidence across multiple scales \cite{song2025understanding,gong2025dark,senden2026evolving}, and population-level analyses provide a natural and robust platform on which different modalities and timescales can be brought together and jointly analyzed. By integrating evidence across scales and allowing them to constrain one another, population neuroscience thus offers a path toward greater integration. For example, when a disorder-associated pattern follows a common axis of population variation \cite{hettwer2026lifespan}, that correspondence alone cannot establish a disorder-specific mechanism. Focused studies can then test candidate explanations by asking whether a targeted perturbation produces the predicted phenotypic change. Furthermore, combined with genetic instruments obtained from large imaging GWAS \cite{elliott2018genome,grasby2020genetic,kumar2025cortical}, Mendelian randomization (MR) \cite{guo2022mendelian,mu2024mendelian,lin2025dissecting} can, under standard assumptions, provide conditional causal inferences along exposure--brain--behavior pathways. For example, two-step MR can be used to test whether imaging-derived phenotypes (IDPs) mediate associations between inflammatory proteins and psychiatric outcomes \cite{zhao2025causal}. Such cross-scale evidence and conditional causal inferences can, on the one hand, provide informative priors and constraints for animal models and interventional studies, and, on the other hand, help subsequent population-level studies to prioritize specific pathways, IDPs, exposures, and high-risk populations.

If the \textbf{Analysis} section emphasized the creation of consensus at multiple levels, then the \textbf{Interpretation} section is where different paths are bridged and connected. Different types of research, including large cohort studies, small deep-phenotyping studies, animal experiments, and computational modeling, can meet and reinforce one another at the interpretive level. In addition to the route already discussed (using population-level priors to guide targeted small-N studies), small studies can themselves be embedded directly within population cohorts. For instance, in a large longitudinal cohort, population-level normative models (as described in the \textbf{Analysis} section) can be used to assign ``coordinates'' to each individual trajectory, indicating along which axes, at what ages, and in what directions and magnitudes an individual begins to deviate from typical trajectories. These ``branching times'' and ``departure directions'' can then be treated as design variables for recruiting specific subgroups into small-N deep studies (e.g., qualitative interviews, targeted task fMRI, pharmacological challenges, or neuromodulation), implementing designs analogous to ``recall-by-brain-pattern'' or ``recall-by-deviation'' (mirroring ``recall-by-genotype'' in genetics \cite{corbin2018formalising,ko2026leveraging}). Concretely, one can first use normative modeling, virtual histology, and brain--behavior axes in population datasets to identify extreme deviators, individuals showing specific deviation patterns, or individuals with atypical timing or rates of developmental change, along certain axes. These individuals can then be selectively recruited into small studies, with repeated baseline measurements or shrinkage estimates helping to account for uncertainty in the deviations guiding selection. Subsequent confirmatory assessments should use separate observations, while intervention effects should be evaluated against contemporaneous controls selected by the same rule, preferably with randomized treatment allocation. In this way, large-N cohorts and small-N deep studies are no longer opposing options but two stages in a continuous pipeline: the population stage calibrates scales, maps the feasible parameter space, and screens potential participants; the small-N stage then probes specific paths, local mechanisms, and intervention sensitivity for selected individuals within that space; finally, findings can be brought back to the population level to test whether the mechanisms generalize broadly or are restricted to particular subgroups.

Within such a workflow, population datasets are not mere background references but instead become active tools for recruitment and stratification, substantially increasing the sampling efficiency and interpretive power of subsequent small-N studies. Conversely, knowledge gained from small studies, through dense longitudinal sampling and tight experimental control, feeds back into population cohort design and is subjected to broader and more systematic testing. For example, if within-individual studies show that a particular network fluctuates on a 1--2 day timescale and is tightly coupled to sleep or mood, later large-scale cohorts can densify sampling at that time scale. If some measures are highly stable within individuals and most variance lies between individuals, designs can prioritize ``many participants, few time points.'' If the opposite holds, ``fewer participants, many time points'' becomes more appropriate. Through such back-and-forth iterations, interpretation not only links scales and research paths, but also provides concrete, actionable guidance for the next round of design and analysis, thereby allowing design--analysis--interpretation to operate as a genuine closed-loop research flywheel.

\section*{A Conceptual Demonstration with Brain Charts}\label{sec4}
As discussed above, only systematically designed population-level studies can fully characterize typical trajectories of human brain development and aging, as well as their variation across sex, ancestry, SES, and geographic and environmental contexts. At this level, the focus is on species-level constraints: general patterns jointly shaped by genes, developmental programs, anatomy, and physical limits. The distributional properties estimated from representative samples then provide the basis for more systematically characterizing inter-individual differences: by computing each individual's deviation from the population at each time point (e.g., a $z$-score or centile score), we can transform brain phenotypes from absolute values into relative values. In this space, deviations provide features for investigating vulnerability, risk, and resilience and, when validated against relevant clinical phenotypes or outcomes, may support disease subtyping and risk stratification in precision psychiatry \cite{marquand2016understanding,marquand2019conceptualizing,chen2021neuroimaging,rutherford2022normative,de2026protocol}. Furthermore, when multi-time-point longitudinal observations are available, centile trajectories provide individualized typical baselines for individuals at different positions in the population distribution. By combining these baselines with intra-individual longitudinal sampling and tracking deviations relative to the corresponding centile trajectories, global age trends are naturally filtered out, and the individual's ``drift'' relative to the group baseline can then be assessed while accounting for measurement error and regression to the mean \cite{barnett2005regression}. Such drift can reflect fluctuations in the individual's state over the relevant time window and offer quasi-experimental leverage and causal interpretive opportunities \cite{liu2021quantifying,bailey2024causal}: time itself provides ordering and direction, making these trajectories well-suited for innovative longitudinal studies of disease course, treatment response, learning, and recovery.

By building interpretable normative models at the population level, population neuroscience integrates these three levels of variation---general patterns, inter-individual differences, and intra-individual longitudinal changes---into a single picture \cite{zhou2024population}. This provides a coherent framework for designing innovative deep-phenotyping studies and for linking and integrating knowledge across multiple spatiotemporal scales. It also provides informative priors for understanding and modeling within-individual development and between-individual differentiation, and for pursuing computational, constructivist theories \cite{astle2023toward} of brain organization. In the context of cross-dataset integration, normative models can further be viewed as a form of ``common currency.'' Different studies may use different protocols and pipelines, making raw values difficult to compare directly. Where the shared normative model is applicable or appropriately calibrated to each study (e.g., through study-specific offsets or transfer learning \cite{wang2025big,gaiser2024estimating,elleaume2026influence}), deviation scores provide a common normative space for comparing results across experiments. In that setting, the normative space itself becomes the shared coordinate system for cross-study integration: instead of asking whether raw brain-phenotype values are directly comparable, we ask how far and in which direction different samples deviate from the relevant normative trajectories under comparable conditions (e.g., age and sex, with variables such as SES included where measurement comparability is supported). Brain charts offer a concrete illustration of this framework: they show how population data can be distilled into a kind of ``population microscope'' that reflects general patterns and provides a baseline for inter-individual differences and intra-individual longitudinal changes, and make it possible to transform data from raw measurement space into a more interpretable normative parameter space.

\begin{figure}[!t]
	\centering
	\includegraphics[width=0.65\columnwidth]{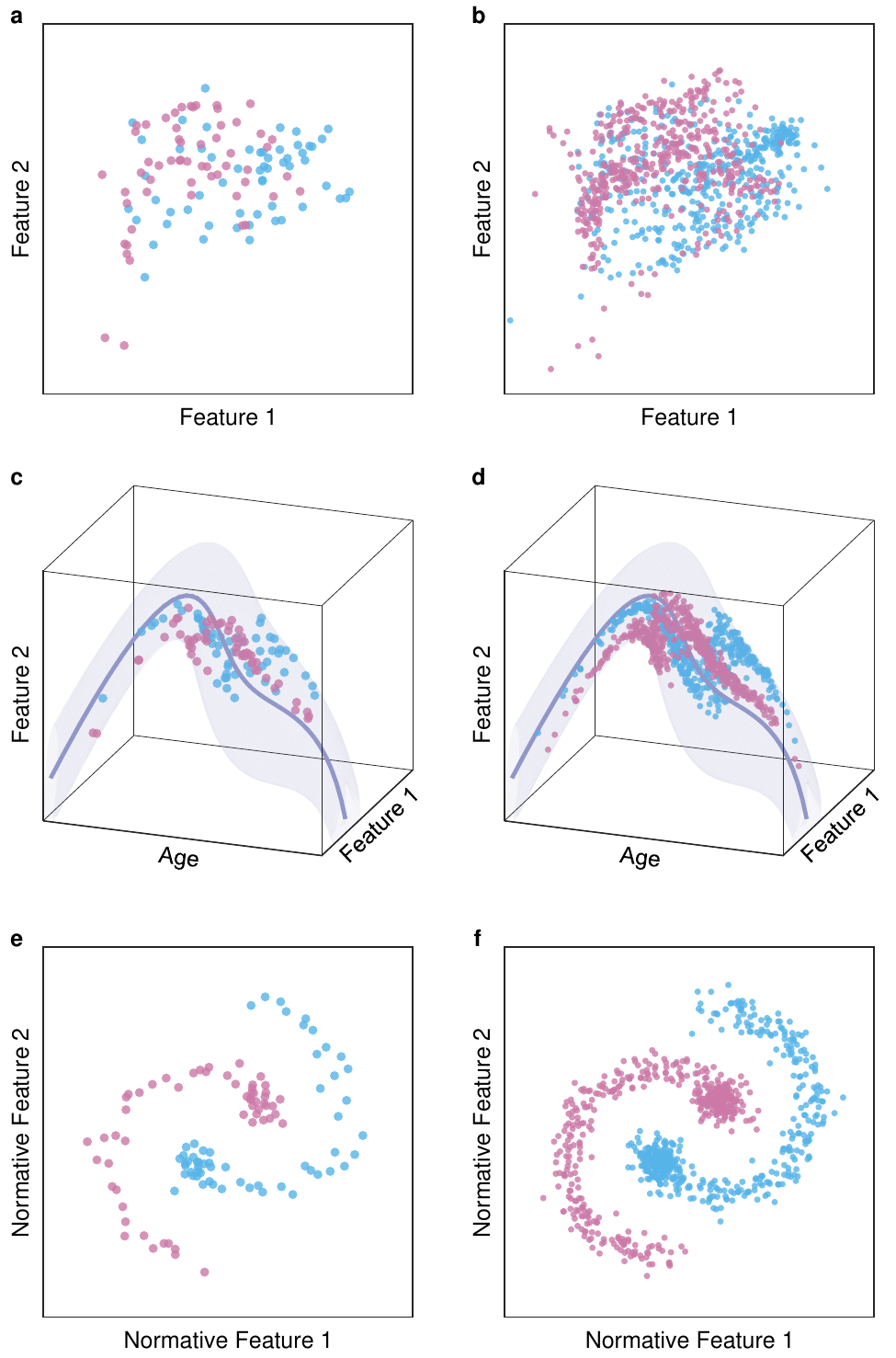}
	\caption{\textbf{From raw measurement space to prior-informed normative coordinates.} \textbf{(a)} The two groups overlap strongly in the raw two-feature projection. \textbf{(b)} When the sample size increases, the respective distribution areas of the two groups appear clearer but still overlap. \textbf{(c)} Adding age as the third dimension shows the observations around the normative developmental trajectory (solid line), while the groups' separation in 3D space is not readily apparent from this viewing angle. \textbf{(d)} Increasing the sample size makes the distribution around the trajectory clearer. \textbf{(e)}~\&~\textbf{(f)} By transforming raw features from measurement space into normative coordinates with reference to the normative trajectory, a previously hidden distinction between the two groups is revealed on the 2D plane defined by the two normative axes, and remains visible even with a relatively small sample size as in panel e. Animations that show the transformation from raw feature scores to normative feature scores, reproducible simulation code, supplementary comparisons of external references with in-sample age adjustment, assessments of type-I error and power, analyses of reference mismatch, local calibration, and reference-estimation uncertainty, and further mathematical perspectives are available at  \href{https://doi.org/10.6084/m9.figshare.30949250}{https://doi.org/10.6084/m9.figshare.30949250}.}
	\label{fig:fig2}
\end{figure}

Figure~\ref{fig:fig2} offers a simplified schematic of one way brain charts can support the framework outlined in this article. In this toy example, only two brain features are measured, so that each participant is represented as a point on a two-dimensional (2D) plane, with different colors indicating different levels of a behavioral trait. The raw features are generated by mapping noisy group-specific latent patterns through nonlinear age trajectories and age-dependent scales that differ above and below the reference trajectory. In panel (a), the distributions of the two color-coded groups overlap heavily in this 2D plane, making it difficult to extract a meaningful brain--behavior association in the original measurement space, even if true latent differences exist. Simply increasing the sample size (panel (b)) makes the contours of each group more clearly defined, but still does not separate the two groups in the 2D projection. Adding age as a third dimension (panels (c) and (d)) shows how the observations are distributed around the normative developmental trajectory (solid line). The groups occupy separate regions in three-dimensional (3D) space, but their separation is not readily apparent in the static projection. A larger sample (panel (d)) makes the distribution around the trajectory clearer. Next, using priors provided by brain charts, we can transform the raw features into deviations from the normative trajectory (panels (e) and (f)). In this 2D plane spanned by normative scores, the two groups that were previously hard to separate become clearly distinguishable; this separation is evident even with relatively small sample sizes (panel (e)). This conceptually illustrates one way small-N studies can benefit from normative priors, and empirical studies increasingly support the potential of analyzing neuroimaging data in a normative space rather than relying only on the raw measurement space \cite{rutherford2023evidence,kobbersmed2026normative,chen2026connectome,liu2026landscaping}. Here, population-level information is brought in as an independently estimated reference distribution. The resulting centile or deviation scores locate individual observations within a shared population frame, providing reusable coordinates for interpreting focused findings and integrating evidence across studies. Modern normative modeling not only estimates age-dependent mean trajectories but can also estimate higher-order statistics, nonlinear age-by-sex effects \cite{gardner2026brain}, and even multivariate joint structure in a latent space, thereby capturing patterns of covariance across brain regions \cite{zhou2023six}. Figure~\ref{fig:fig2} illustrates only the simplest case. In practice, the more consequential transformations often take place in high-dimensional, nonlinear representational spaces, which may need to be learned using manifold-based or deep generative approaches \cite{attye2024data,you2025conditional,lawry2025deep}. An important future direction is therefore to systematically build, compare, and iteratively refine such transform spaces, so as to better support individualized, small-sample studies and move toward more precise mechanistic understanding and translational applications.

\section*{Conclusions}\label{sec5}
Shaped by waves of reproducibility crises \cite{marek2022reproducible,zuo2014open,botvinik2020variability} and the recent surge of big data, human neuroscience has entered a period of both unprecedented opportunity and deepening fragmentation. On the one hand, large-scale consortium studies and population-level BWAS have flourished; on the other hand, consensus has fractured around basic assumptions, priorities in study design, analytic approaches, and interpretive frameworks. In this article, we argue that population neuroscience offers a more integrative picture: BWAS, brain charts, and related population-level agendas can be viewed not merely as big-N exercises in finding significant associations, but as a long-term, systematic slow-science effort to build reference and interpretive frameworks that span disciplines, modalities, and spatiotemporal scales. Within this framework, ontogeny (development) occupies a central axis. On the one hand, it carries species-level constraints shaped by function and evolution (e.g., genetically and developmentally constrained trajectories); on the other, it provides the temporal substrate on which proximate mechanisms, such as network plasticity, learning, and adaptation, are directly observed in dense longitudinal and interventional studies. Observing and deciphering how genetic and environmental influences shape diverse developmental trajectories at the population level is a necessary step toward understanding the brain's variability and complexity. Designs that are cross-organ, cross-generational, and cross-cultural embed individual differences within broader biological systems, life-course processes, and ecological contexts, helping us address the core question that motivated population neuroscience from the outset: \textit{why do human brains differ?}

If reproducible BWAS by Marek \textit{et al.} have highlighted the necessity of ``detecting weak signals'' in large samples, then the next step advocated by population neuroscience is to ``build shared priors'' in large samples and to treat these priors as reference infrastructure. Within the design--analysis--interpretation cycle, population references guide focused mechanistic studies, whose findings generate hypotheses for independent population-level evaluation and priorities for subsequent study design. Brain charts are only an opening move in this agenda, demonstrating how population data can be used to construct coordinate systems that simultaneously serve general patterns, individual differences, and longitudinal changes. A broader vision is that future neuroscience may come to be organized less as a loose collection of disconnected papers, and more as a set of continually upgraded reference resources, including maps, models, weight libraries, and simulation priors, together with a body of mechanistic studies and clinical applications that operate on top of them \cite{yarkoni2010cognitive}. In this way, evidence accumulated across spatiotemporal scales, populations, and subgroups can be brought into alignment, gradually shifting the field from a patchwork of local stories toward a more coherent, iteratively updated narrative.

If the past three decades of in vivo human neuroimaging have revealed the complexity and variability of the brain, then the next three decades of population neuroscience may place the field in a stronger position to address more fundamental questions: where do individual differences come from? Which variations are largely inevitable products of genes, development, and environment, and which are plastic and reversible? Population-level evidence offers one of the most promising lenses through which to understand why brains differ from one another, and on that basis construct mechanistic models and intervention strategies that both respect individual uniqueness and retain population-level generalizability, thereby informing clinical practice, education, and public health policy.

\section*{Competing Interests}
The authors declare no conflict of interest.

\section*{Acknowledgements}
We thank the \textit{Research Program on Discipline Direction Prediction and Technology Roadmap} of China Association for Science and Technology for bibliometric resources, the \textit{Chinese Color Nest Data Community} (\url{https://ccndc.scidb.cn/en}) at National Basic Science Data Center and the \textit{Lifespan Brain Chart Consortium} (\url{https://github.com/brainchart/lifespan}) for big data resources. The authors used ChatGPT to assist with language editing and polishing. The authors reviewed and approved all revisions and take full responsibility for the content of the manuscript.

\section*{Funding}
This study has been supported by the Brain Science and Brain-like Intelligence Technology - National Science and Technology Major Project (2021ZD0200500): The Chinese Child Brain Development (CCBD) project.

\section*{Data Availability}
All data used in the conceptual demonstration and supplementary analyses are simulated and generated by the deposited code. The MATLAB visualization script, executed Python notebook and HTML export, static figure, animations, and supplementary mathematical discussion are available at \href{https://doi.org/10.6084/m9.figshare.30949250}{https://doi.org/10.6084/m9.figshare.30949250}.

\bibliographystyle{unsrt}
\bibliography{manuscript}  

\end{document}